\documentclass[10pt]{iopart}
\usepackage{iopams}
\usepackage{graphicx}
\usepackage{setstack}
\eqnobysec

\newcommand{\bk}{\bi{k}}

\newcommand{\adkisig}{a^{\dagger}_{\bk i \sigma}}
\newcommand{\akisig}{a^{\phantom\dagger}_{\bk i \sigma}}
\newcommand{\akpisig}{a^{\phantom\dagger}_{\bk' i \sigma}}

\newcommand{\lmtwo}{LM$^{SU(2)}$}
\newcommand{\lmfour}{LM$^{SU(4)}$}

\newcommand{\tkfour}{T_K^{SU(4)}}

\newcommand{\di}{D_{i}(\omega)}
\newcommand{\gi}{G_{ii}(\omega)}

\begin{document}
\title{Dynamics of capacitively coupled double quantum dots.}
\author{Martin R Galpin\dag, David E Logan\dag\ and H R Krishnamurthy\ddag}
\address{\dag\ Oxford University, Physical and Theoretical Chemistry
Laboratory, South   
Parks Road, Oxford OX1 3QZ, UK.}
\address{\ddag\ Department of Physics, IISc, Bangalore 560 012, and
JNCASR, Jakkur, Bangalore, India.}
\begin{abstract}
We consider a double dot system of equivalent, capacitively coupled semiconducting quantum dots, each coupled to its own lead, in a regime where there are two electrons on the double dot. Employing the numerical renormalization group, we focus 
here on single-particle dynamics and the zero-bias conductance, considering in particular the rich range of behaviour arising as the interdot coupling
is progressively increased through the strong coupling (SC) phase, from
the spin-Kondo regime, across the $SU(4)$ point to the charge-Kondo regime; 
and then towards and through the quantum phase transition to a charge-ordered (CO) phase. We first consider the two-self-energy description required to describe the broken symmetry CO phase, and implications thereof for the non-Fermi liquid nature of this phase. Numerical results for single-particle dynamics on all frequency scales are then considered, with particular emphasis on universality and scaling of low-energy dynamics throughout the SC phase. The role of symmetry breaking perturbations is also briefly discussed.
\end{abstract}
\pacs{71.27.+a, 72.15.Qm, 73.63.Kv}

\vspace{28pt plus 10pt minus 18pt} \noindent{\small\rm Published as: Martin R Galpin et al 2006 {\it J. Phys.: Condens. Matter} {\bf 18} 6571-6583\\ Journal of
Physics: Condensed Matter \copyright\ 2006 IOP Publishing Ltd. \par}

\maketitle
\section{Introduction}
\label{sec:intro}
In a previous paper~\cite{I} (referred to hereafter as I) we have studied a symmetrical, semiconducting double quantum dot system, with interdot capacitive coupling in addition to the usual intradot Coulomb interaction. Employing the numerical renormalization group (NRG) approach~\cite{wilson,kww1,kww2,hewson}, and focussing on the regime with two electrons on the double dot, 
a rich and diverse range of physical behaviour was shown to arise on increasing the 
ratio $U'/U$ of interdot to intradot coupling strengths~\cite{ddprl}.
The system first evolves continuously from a spin-Kondo state, where the dot spins are in essence separately quenched ($SU(2) \times SU(2)$), to an $SU(4)$ Kondo state with entangled charge and spin degrees of freedom arising at $U'/U =1$. This in turn
precedes a rapid but smooth crossover to a charge-Kondo state in which a charge-pseudospin is Kondo quenched; followed by the suppression of charge-pseudospin tunneling, a continuous collapse of the underlying low-energy Kondo scale, and a Kosterlitz-Thouless quantum phase transition at a critical $U'_{c}$ to a degenerate charge-ordered state, itself a non-Fermi liquid with a $\ln 2$ residual entropy.

  Our primary emphasis in I was on the structure, stability and flows between
the underlying RG fixed points, on the overall phase diagram and evolution of the characteristic low-energy Kondo scale arising for all $U' < U'_{c}$ in the strong coupling phase; and on static physical properties such as spin- and charge-susceptibilities, including some exact results for associated Wilson ratios. In the present paper by contrast we focus on dynamical properties of the system, specifically the $\omega$-dependence and $U'$-evolution of local single-particle spectra, which at the Fermi level in particular determine the ($T=0$) linear differential conductance across one or other dot (the interdot coupling is purely capacitive, so there is of course no transport through the pair of dots).

  A number of theoretical issues are discussed in section~\ref{sec:theory}. These relate in particular to the `two-self-energy' description that we show is required 
to describe the broken symmetry charge-ordered (CO) phase, its connection both to the potential scattering inherent to the CO phase and to the `single' self-energy of
conventional field theory; and hence insights into the non-Fermi liquid character of the CO phase. Numerical results for dynamics are presented in section~\ref{sec:numresdyn}, including the important issues of universality and scaling of the low-energy single-particle spectra throughout the strong coupling phase.

\section{Theoretical issues}
\label{sec:theory}

  We focus then on the local single-particle spectrum for dot $i$,
$\di = -\case{1}{\pi} \mathrm{Im}\gi$, with $\gi$ the Fourier transform of the (retarded) single-particle propagator $G_{ii}(t) = -i\theta (t)\langle\{c_{i\sigma}(t),c^{\dagger}_{i\sigma}\}\rangle$. In terms of the local spectrum, the zero-bias differential conductance across dot $i$ at $T=0$, here denoted by $g_{i}$, is given by \cite{meirwin}
\begin{equation}
g_i=\left(\frac{\partial J_i}{\partial V_i}\right)_{V_i=0}=\left(\frac{2e^2}{h}\right)\pi\Gamma D_i(\omega=0)
\end{equation}
with $J_i$ and $V_i$ respectively the current and voltage across the dot (and
$\Gamma =\pi V^2 \rho$ the hybridization strength, as in I). The $T=0$ linear differential conductance thus probes the single-particle spectrum at the Fermi level, $\omega =0$ (and at finite, low bias voltage $V_{\mathrm{sd}}$, 
$\pi\Gamma D_{i}(\omega = eV_{\mathrm{sd}})$ provides an approximation to the
conductance).
More generally, from the discussion given in I, one might expect the $\omega$-dependence of $\di$ to exhibit a rich evolution as the interdot interaction $U'$ is progressively increased through the spin-Kondo regime ($U' < U$), across the $SU(4)$ point $U'=U$ into the charge-Kondo regime; and then towards the quantum phase transition occurring at $U'_{c}$ and into the broken symmetry charge ordered (CO) phase.

\begin{figure}
\centering\includegraphics{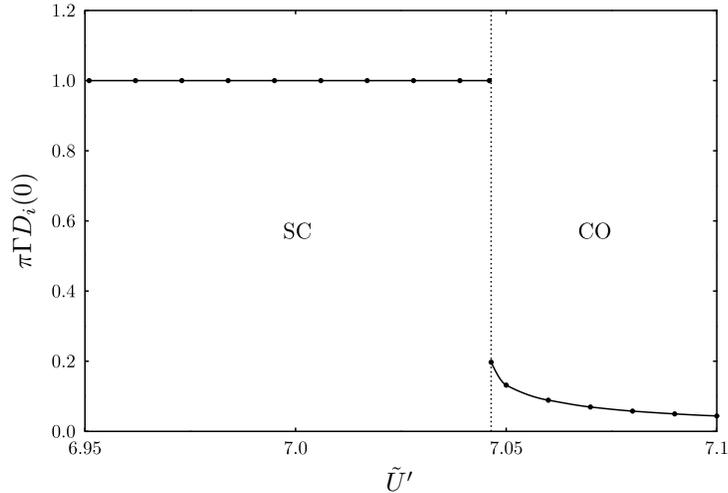}
\caption{\label{fig:specfermi} NRG results for the zero-frequency spectral density $\pi\Gamma D_i(\omega=0)$, or equivalently the linear differential conductance $g_{i}/(2e^{2}/h)$, as a function of $\tilde U'$ for fixed $\tilde U=7$. The dotted line marks the phase boundary $\tilde U'_c\simeq 7.046$.} 
\end{figure}

  To motivate subsequent discussion, we shown in figure~\ref{fig:specfermi}
NRG results for
the linear differential conductance $g_{i}/(2e^2/h) = \pi \Gamma D_{i}(\omega =0)$
\emph{vs} the interdot $\tilde{U}' = U'/(\pi \Gamma)$ for a fixed intradot
$\tilde{U} = U/(\pi \Gamma) = 7$. As discussed further below, the transition between the strong coupling (SC) and CO phases (here occurring at $\tilde{U}'_{c} \simeq 7.046$) is clearly evident: throughout the SC phase the conductance is `pinned' at the unitarity limit of $g_{i} = 2e^2/h$, while it drops discontinuously on entering the CO phase and decreases montonically thereafter with increasing 
$\tilde{U}'$.

\subsection{Two-self-energy description}

 In considering single-particle dynamics a number of rather subtle issues arise,
which as we show have important ramifications e.g.\ in understanding the non-Fermi liquid nature of the CO phase. The first concerns the `two-self-energy' description that is required to describe the CO phase.

The SC phase is a Fermi liquid state, adiabatically connected to the non-interacting limit $U=0=U'$ and with Kondo screening ensuring that the ground state is locally non-degenerate. As such
the traditional `single self-energy' description,
based ultimately on perturbation theory in the interaction, is appropriate;
where (with $\omega^{+} = \omega + i0+$)  $\gi$ is expressed as
\begin{equation}
\label{eq:sse}
G_{ii}(\omega) = \frac{1}{[\omega^{+} - \epsilon + i\Gamma - \Sigma_{i} (\omega) ]}
\end{equation}
in terms of the conventional self-energy $\Sigma_{i} (\omega)$ ($ = \Sigma_{i}^{R}(\omega) - i\Sigma_{i}^{I}(\omega)$ with the imaginary part $\Sigma_{i}^{I}(\omega) \ge 0$).

The CO phase by contrast is not adiabatically connected to the non-interacting limit
(and as such is a non-Fermi liquid), as evident e.g.\ from the phase diagram figure 4 of I, where the CO phase arises for $U' > U'_{c}$ and the critical $U'_{c}$ as a function of $U$ satisfies $U'_{c}(U) > 0$ for all $U \ge 0$.
This phase is of course characterised by charge symmetry-breaking, the ground state being \emph{doubly} degenerate ($(n_L,n_R) = (2,0)$ and $(0,2)$ occurring with equal 
weight in the fixed point Hamiltonian, section 4.1 of I). For this reason, as is  
obvious from its Lehmann representation, the local propagator in the CO phase is 
of form
\begin{equation}
\label{eq:tse1}
G_{ii}(\omega) = \case{1}{2} [G_{ii;A}(\omega) + G_{ii;B}(\omega)].
\end{equation}
Here $G_{ii;\alpha}(\omega)$, with corresponding spectral density $D_{i\alpha}(\omega)=-\case{1}{\pi}\mathrm{Im}G_{ii;\alpha}(\omega)$,
denotes the propagator obtained from one or other of the degenerate ground states $\alpha = A$ or $B$ mentioned above. This in turn may be expressed as
\begin{equation}
\label{eq:tse2}
G_{ii;\alpha}(\omega) = \frac{1}{[\omega^{+} - \epsilon + i\Gamma - \tilde{\Sigma}_{i\alpha}(\omega) ]}
\end{equation}
in terms of a self-energy $\tilde{\Sigma}_{i\alpha}(\omega)$ ($=\tilde{\Sigma}^{R}_{i\alpha}(\omega) -i\tilde{\Sigma}^{I}_{i\alpha}(\omega)$). 
The two-self-energy (TSE) description embodied in equations 
(\ref{eq:tse1}, \ref{eq:tse2})
is a necessity and not a luxury when considering the CO phase (although equation
(\ref{eq:sse}) may obviously still be used to \emph{define} a single-self-energy in the CO phase, as considered in section 2.3 below). If one sought for example to calculate $\gi$ diagrammatically in the CO phase, one would construct self-energies from \emph{charge} symmetry-broken mean-field propagators (unrestricted Hartree Fock); the double degeneracy of the mean-field saddle points 
ensuring that it is the $\tilde{\Sigma}_{i\alpha}(\omega)$ and hence $G_{ii;\alpha}(\omega)$ that are thereby calculated, with the full $\gi$ then obtained from
(\ref{eq:tse1}).

  Notice also that, trivially, in a case where $\tilde{\Sigma}_{i\alpha}(\omega)
\equiv \Sigma_{i}(\omega)$ is independent of $\alpha$, the TSE description reduces
formally to the conventional single self-energy description of equation
(\ref{eq:sse}). It thus encompasses the SC phase simply as a special case.\\

  We now consider the implications of a TSE description, considering primarily the CO phase. The first point to note here is that, at the Fermi level $\omega =0$, the imaginary part of the self-energies $\tilde{\Sigma}_{i\alpha}(\omega)$ vanishes,
\begin{equation}
\label{eq:imsig}
\tilde{\Sigma}^{I}_{i\alpha}(\omega =0) = 0
\end{equation}
for $\alpha =A,B$. This reflects the fact that the fixed point Hamiltonian for the CO phase, section 4.1 of I, is characterised purely by potential scattering of the conduction electrons (a point pursued in section 2.2 below). It can also be shown by considering the diagrams for $\tilde{\Sigma}_{i\alpha}(\omega)$ constructed from charge symmetry-broken mean-field propagators. From equations (\ref{eq:tse2}, \ref{eq:imsig}) it follows directly that the Fermi level spectral density is given by
\begin{equation}
\label{eq:dialpha}
\pi \Gamma D_{i\alpha}(\omega =0) = \frac{1}{[{\epsilon_{i\alpha}}/{\Gamma}]^{2} +1}
\end{equation}
where $\epsilon_{i\alpha}$ denotes the `renormalized level'
\begin{equation}
\label{eq:renormlevel}
\epsilon_{i\alpha} = \epsilon + \tilde{\Sigma}^{R}_{i\alpha}(\omega =0)
\end{equation}
(satisfying the obvious `left/right' symmetry
$\epsilon_{L\alpha} = \epsilon_{R\bar{\alpha}}$ with $\bar{\alpha} = B$ or $A$ for
$\alpha = A$ or $B$ respectively);
and such that the linear differential conductance 
\begin{equation}
\label{eq:ldc1}
\pi \Gamma D_{i}(0) =
\case{1}{2}\sum_{\alpha =A,B} (\pi \Gamma D_{i\alpha}(0)).
\end{equation}
 This result may be recast equivalently in terms of the conduction electron
phase shift $\delta_{i\alpha} = \delta_{i\alpha}(\omega =0)$, where 
$\delta_{i\alpha}(\omega) = \mathrm{arg}(T_{i\alpha}(\omega))$ with
$T_{i\alpha}(\omega) = V^{2}G_{ii;\alpha}(\omega)$ the conduction electron
t-matrix. Using equations (\ref{eq:tse2}, \ref{eq:imsig}) gives
$\delta_{i\alpha} = \mathrm{arctan}(\case{\Gamma}{\epsilon_{i\alpha}})$ and hence
$\pi \Gamma D_{i\alpha}(\omega =0) = \mathrm{sin}^{2}(\delta_{i\alpha})$, i.e.\
\begin{equation}
\label{eq:ldcphaseshifts}
\pi \Gamma D_{i}(\omega =0) = \case{1}{2}\sum_{\alpha = A,B} \mathrm{sin}^{2}(\delta_{i\alpha})
\end{equation}
for the linear differential conductance.

  The considerations above relating the Fermi level spectrum to the renormalized levels and/or phase shifts
are quite general. For the particular case of particle-hole symmetry ($\epsilon = -\case{U}{2} -U'$, for which the explicit NRG calculations we show here are performed) one has the additional, intuitively obvious reflection symmetry about the Fermi level, $D_{iA}(\omega) = D_{iB}(-\omega)$. For the self-energies this corresponds to 
$\tilde{\Sigma}^{I}_{iA}(\omega) = \tilde{\Sigma}^{I}_{iB}(-\omega)$ and
\begin{equation}
\label{eq:phsymm}
\epsilon + \tilde{\Sigma}^{R}_{iA}(\omega) = -[\epsilon + 
\tilde{\Sigma}^{R}_{iB}(-\omega)].
\end{equation}
The renormalized levels (equation (\ref{eq:renormlevel})) thus satisfy
\begin{equation}
\epsilon_{iA} = -\epsilon_{iB}
\end{equation}
implying equal and opposite scattering phase shifts, 
$\delta_{iA} = -\delta_{iB}$ (as noted in section 4.1 of I), such that the linear differential conductance 
$\pi \Gamma D_{i}(0) \equiv \mathrm{sin}^{2}(\delta_{i\alpha})$.

  As already noted the SC phase is recovered simply as a limit of the above analysis, with $\tilde{\Sigma}_{i\alpha}(\omega) \equiv \Sigma_{i}(\omega)$ independent of 
$\alpha$. At particle-hole symmetry in this case, equations 
(\ref{eq:renormlevel}, \ref{eq:phsymm}) show that the renormalized level \emph{vanishes} by symmetry (corresponding to a phase shift $\delta = \case{\pi}{2}$), and hence from equations (\ref{eq:dialpha}) or (\ref{eq:phsymm}) that
$\pi \Gamma D_{i}(0) = 1$ -- the unitarity limit, seen clearly in the NRG results
of figure~\ref{fig:specfermi} for $U' < U$ in the SC phase.

\subsection{Potential scattering}
\label{sec:potscat}

  We now relate the above results for single-particle dynamics at the Fermi level
to potential scattering. To this end consider the Hamiltonian
\begin{equation}
\label{eq:PS}
H_i^{PS}=\sum_{\bk,\sigma}\epsilon_\bk\adkisig\akisig+
K_{i\alpha}\sum_{\bk,\bk',\sigma}\adkisig\akpisig
\end{equation}
in which potential scattering is operative in conduction channel $i$ (with strength
$K_{i\alpha}$), and for which the conduction electron t-matrix  
$T_{i\alpha}(\omega)$ is readily calculated~\cite{hewson}. Equating the resultant 
$T_{i\alpha}(0)$ to $T_{i\alpha}(0) = V^{2}G_{ii;\alpha}(0)$ considered above, 
one finds that
(a) $\tilde{\Sigma}^{I}_{i\alpha}(0) =0$ , i.e.\ equation (\ref{eq:imsig})
is recovered;
and (b) that the renormalized level $\epsilon_{i\alpha}$ is related to the scattering strength by
\begin{equation}
\label{eq:wibbly}
\frac{\epsilon_{i\alpha}}{\Gamma} = \frac{-1}{\pi \rho K_{i\alpha}}
\end{equation}
(with $\rho$ as usual the conduction band density of states). 

Equation
(\ref{eq:PS}) is in essence the fixed point Hamiltonian for the CO phase.
More precisely, referring to equation (4.2) of I and noting that $f^{\dagger}_{0i\sigma} \propto \sum_{\bk}\adkisig$, the fixed point Hamiltonian is of form
$\sum_{i} H_{i}^{PS}$ (both channels are of course involved); with
$K_{i\alpha} \equiv (n_{i}-1)K$ for the particle-hole symmetric case considered explicitly, where the dot charges $(n_{L},n_{R}) = (0,2)$ or $(2,0)$
correspond respectively to the broken symmetry states we have denoted by $\alpha = A$ or $B$. Hence, considering explicitly the $i = L$ channel, 
\begin{equation}
\label{eq:caramel}
\frac{\epsilon_{iA}}{\Gamma} = \frac{1}{\pi \rho K} = \frac{-\epsilon_{iB}}{\Gamma}
\end{equation}
with corresponding phase shifts $\delta_{iA} = \mathrm{arctan}(\pi \rho K)
= -\delta_{iB}$, such that from equation (\ref{eq:ldcphaseshifts}) the linear conductance is given by
\begin{equation}
\label{eq:ldcfrac}
\pi \Gamma D_{i}(0) = \frac{1}{1 + 1/(\pi \rho K)^{2}}
\end{equation}
in terms of the potential scattering strength $\rho K$ (and is obviously independent of whether $i =L$ or $R$ is considered). It is straightforward to relate 
$\rho K$ in turn to the potential scattering coupling constant $\tilde{K}$ that enters
the $\Lambda$-discretized NRG Hamiltonian for the CO fixed point (equation (4.3) of I): 
specifically, 
$\rho K = ([1-\Lambda^{-1}]/[2\ln\Lambda]) \tilde{K}$.
Since $\tilde{K}$ itself is directly determined numerically in the NRG calculations,
the linear conductance can thus either be determined this way from equation
(\ref{eq:ldcfrac}), or as the $\omega =0$ limit of the single-particle spectrum
$\di$ (determined via the standard NRG method~\cite{hewsoncosti} reprised at the 
beginning of section 3). In practice we find the two methods to be in very good numerical agreement.

  As detailed in section 4.1 of I the CO phase corresponds to a \emph{line} of 
fixed points, one for each $U' > U'_{c}$. The critical endpoint, occurring at 
$U' =U'_{c}+$ on the CO side of the  transition, corresponds to a critical 
$\tilde{K}_{c}$ given by equation (4.8) of I, and thus to a critical 
$\pi \rho K_{c} = \mathrm{tan}[\case{\pi}{2}(1-\case{1}{\sqrt{2}})]$.
From equation (\ref{eq:ldcfrac}) the differential conductance
$g_{i}/(2e^{2}/h)$, which is unity throughout the SC phase for $U' < U'_{c}$,
thus drops abruptly at the transition to a value of 
$\simeq 0.197..$ at $U' = U'_{c}+$~\cite{garst};
as seen clearly in the NRG results of figure~\ref{fig:specfermi}. This 
discontinuous drop in the differential conductance as the transition is crossed 
appears to be a rather general signature of a Kosterlitz-Thouless quantum phase transition, it being found also for a multi-level small dot close to a singlet-triplet degeneracy point~\cite{HofSch} and for a pair of Ising-coupled Kondo impurities, onto which maps the problem of spinless, capacitively coupled metallic islands close to the degeneracy point between $N$- and $N+1$-electron states~\cite{garst}.

  We also add that while the existence of an abrupt drop in the differential conductance as the transition is crossed is not specific to the particle-hole symmetric case shown in figure~\ref{fig:specfermi}, the magnitude of the discontinuity is, reflecting the fact that deviation from particle-hole symmetry generates additional potential scattering of the \emph{same} sign on the two leads.
The generic form for the linear conductance follows from equations 
(\ref{eq:dialpha}, \ref{eq:ldc1}, \ref{eq:wibbly}) as
\begin{equation}
\pi \Gamma D_{i}(0) = \frac{1}{2}\biggl\{\frac{1}{1 + 1/(\pi \rho K_{iA})^
{2}}~+~ \frac{1}{1 + 1/(\pi \rho K_{iB})^{2}} \biggr\}
\end{equation}
with the $K_{i\alpha}$ of form (again for $i=L$ explicitly) $K_{iA} = -K+\delta K$ and $K_{iB} = K+\delta K$,
where $\delta K$ represents the additional potential scattering common to each channel
that vanishes only at particle-hole symmetry.

\subsection{Non-Fermi liquid behaviour}

The conventional description of single-particle dynamics centres 
on the usual single self-energy $\Sigma_{i}(\omega)$, defined by equation (\ref{eq:sse}) (and the Dyson equation implicit therein). The analysis above has however been couched in terms of a TSE description, so the obvious question arises: what are the implications for $\Sigma_{i}(\omega)$?

  The general relation between $\Sigma_{i}(\omega)$ and the two self-energies
$\tilde{\Sigma}_{i\alpha}(\omega)$ follows simply from direct comparison of equations (\ref{eq:sse}) with (\ref{eq:tse1}, \ref{eq:tse2}), and is given by
\begin{equation}
\label{eq:ssetse}
\fl \Sigma_{i}(\omega) = \case{1}{2}[\tilde{\Sigma}_{iA}(\omega) + \tilde{\Sigma}_{iB}(\omega)] ~~
+ ~~ \frac{[\case{1}{2}(\tilde{\Sigma}_{iA}(\omega) - \tilde{\Sigma}_{iB}(\omega))]^{2}}{\omega^{+} +i\Gamma - [\epsilon + \case{1}{2}(\tilde{\Sigma}_{iA}(\omega) + \tilde{\Sigma}_{iB}(\omega))]}.
\end{equation}
Since $\tilde{\Sigma}^{I}_{i\alpha}(\omega =0) =0$ (equation (\ref{eq:imsig})), 
the imaginary part of the single self-energy at the Fermi level follows from
equation (\ref{eq:ssetse}) as
\begin{equation}
\label{eq:ssefl}
\Sigma_{i}^{I}(\omega =0)~ = ~ \frac{[\epsilon_{iA} -\epsilon_{iB}]^{2}\Gamma}{[\epsilon_{iA}+\epsilon_{iB}]^{2} + 4\Gamma^{2}} 
\end{equation}
in terms of the renormalized levels $\epsilon_{i\alpha}$, equation 
(\ref{eq:renormlevel}).
For the SC phase, where $\epsilon_{iA} = \epsilon_{iB}$ generically (and
$\epsilon_{iA} =0 =\epsilon_{iB}$ at particle-hole symmetry), equation 
(\ref{eq:ssefl})
gives $\Sigma_{i}^{I}(\omega =0) =0$ --- just as required for a Fermi liquid state 
($\Sigma_{i}^{I}(\omega) \propto \omega^{2}$ as $\omega \rightarrow 0$).

  For the degenerate CO phase by contrast, the renormalized levels $\epsilon_{iA} \neq \epsilon_{iB}$. From equation (\ref{eq:ssefl}), the conventional self-energy thus has a non-vanishing imaginary part at the Fermi level $\omega =0$. This is of course a direct reflection of the non-Fermi liquid nature of the CO phase; and we emphasise that to recover this behaviour necessitates use of the two-self-energies $\tilde{\Sigma}_{i\alpha}(\omega)$ (which themselves satisfy $\tilde{\Sigma}_{i\alpha}^{I}(\omega =0) =0$). For the specific case of particle-hole symmetry, 
$\epsilon_{iA} =-\epsilon_{iB}$ and
\begin{equation}
\frac{\Sigma_{i}^{I}(\omega =0)}{\Gamma} = \biggl[\frac{\epsilon_{i\alpha}}{\Gamma}\biggr]^{2} ~
\equiv ~ \frac{1}{(\pi \rho K)^{2}}
\end{equation}
(where equation (\ref{eq:caramel}) relating $\epsilon_{i\alpha}$ to the potential scattering strength is also used). And at $U' = U'_{c}+$, using the critical 
$\pi \rho K_{c}$, $\Sigma_{i}^{I}(0)$ is then given explicitly by
\begin{equation}
\frac{\Sigma_{i}^{I}(\omega =0)}{\Gamma} = \mathrm{cot}^{2}[\case{\pi}{2}(1-\case{1}{\sqrt{2}})]~~\simeq 4.1~~~~~~~~~~~:U' =U'_{c}+
\end{equation}

\section{Numerical results: dynamics}
\label{sec:numresdyn}

\begin{figure}
\centering\includegraphics{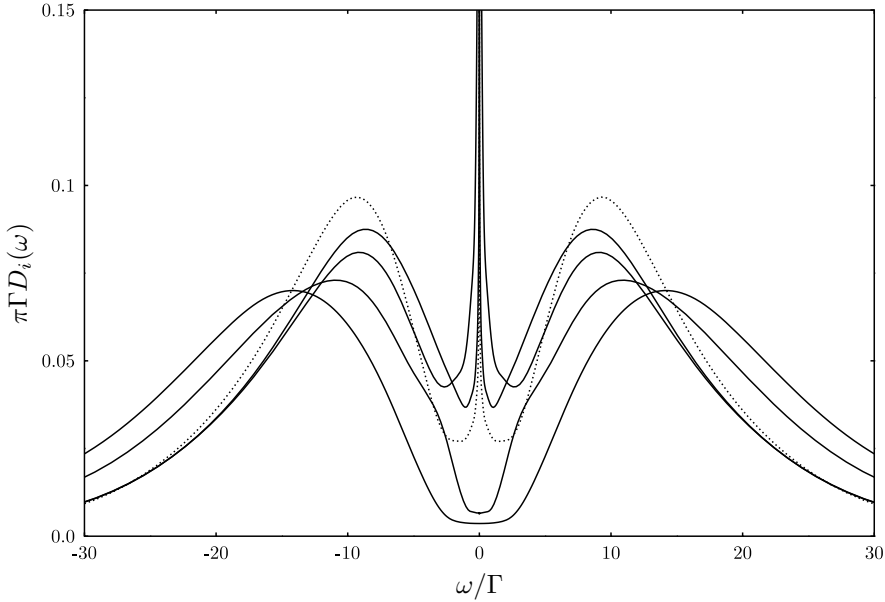}
\caption{\label{fig:hubbsat} Single-particle spectra on `high' frequency scales:
$\pi \Gamma D_{i}(\omega)$ \emph{vs} $\omega/\Gamma$ for $\tilde U'=0$ (dotted line), $6$, $7$, $8$ and $9$ (solid lines, top to bottom) with a fixed $\tilde U=7$.}
\end{figure}

We now consider NRG results for the full frequency dependence of the local dynamics,
analysing spectra (for a range of bare parameters) on both high and low frequency scales, and in particular (section~\ref{sec:uniscale})
extracting the universal scaling behaviour in the SC phase. 
The single-particle spectrum at finite-$\omega$ is obtained using the standard method \cite{hewsoncosti} --- a set of poles corresponding to the single-particle excitations are calculated from the sequence of NRG iterations, and these are subsequently broadened on a logarithmic scale to compensate for the initial logarithmic discretization of the Hamiltonian and recover the continuum.

\subsection{Spectra on high $\omega$ scales}
\label{sec:hubb}
We begin by considering briefly the evolution of single-particle dynamics on non-universal `high' energy scales --- in particular the Hubbard satellites --- upon progressively increasing the interdot interaction $U'$ from the limit $U'=0$ where the dots are uncoupled. Figure~\ref{fig:hubbsat} shows
$\pi\Gamma D_i(\omega)$ versus $\omega/\Gamma$ for a fixed $\tilde U=7$ and $\tilde U'=0$ (dotted line), $6$, $7$, $8$ and $9$ (solid lines, top to bottom at $\omega/\Gamma=10$).
For all $\tilde U'\le\tilde U$, the Hubbard satellites are seen to be centred at approximately the same frequency, whereas for $\tilde U'>\tilde U$ the positions of the satellites become $\tilde U'$-dependent. 

  It is straightforward to explain this behaviour by recourse to the atomic limit $\Gamma=0$. For $U'<U$, the atomic limit ground state is $(n_{L},n_{R}) = (1,1)$ (see equation (2.3) of I). This connects via single-particle excitations (for e.g.\ the left dot) to $(0,1)$ and $(2,1)$ which, regardless of $U'$, are both an energy 
$U/2$ higher than the ground state. The atomic limit single-particle spectra thus 
exhibit Hubbard satellites (poles in this case) at 
$\omega = \pm \case{U}{2}$, independently of $U'$, which is the essential origin of the behaviour seen in figure~\ref{fig:hubbsat} for $U' < U$.
For $U'>U$ by contrast, the atomic
limit ground state is $(2,0)$/$(0,2)$. The corresponding excited states accessible under single-particle excitations, $(1,0)$ and $(1,2)$, are now an energy $U'-U/2$ higher and thus become increasingly separated from the ground states as $\tilde U'$ is raised; generating $U'$-dependent Hubbard satellites at 
$\omega = \pm (U'-U/2)$, as again seen in figure~\ref{fig:hubbsat}.\\

  Before considering dynamics on the key low-energy scales, we make a
general remark. NRG spectra may of course be calculated from either the even or the odd set of iterations (fixed points are inexorably fixed points of the \emph{square} of the RG transformation~\cite{wilson,kww1}); and the same spectrum $\di$ naturally 
results in either case. For the CO phase arising for $U' > U'_{c}$ the NRG ground state is as one would expect doubly degenerate, regardless of whether even or odd iterations are considered. The two-self-energy description considered above is thus ubiquitous in this case. For $U' < U'_{c}$ in the SC phase by contrast there is an important difference between even-$N$ and odd-$N$ iterations,
just as found in~\cite{JPCM} for the single-impurity Anderson model.
In the even-$N$ case the NRG ground state is always (for any iteration) 
non-degenerate, for which 
reason an effective single-self-energy description of dynamics obviously arises.
 But for the odd-$N$ iterations the NRG ground state is always found
to be doubly degenerate (and with `overlap' on the dots). This leads directly
to an inherent two-self-energy description of local dynamics, even in this 
non-symmetry-broken SC phase.
We emphasise again that, whether even or odd iterations are considered, the resultant $\di$ obtained from the NRG calculations is the same.
But the key point is that a two-self-energy description of  dynamics, which is a necessity in describing the broken symmetry CO phase, is equally applicable in describing the SC phase; a point not widely appreciated, but one which also underlies the Local Moment Approach to correlated electron systems~\cite{lma1,lmamtg,lmasg1}.
The situation sketched here is directly parallel to that detailed in~\cite{JPCM}, 
to which the reader is referred for further discussion.

\subsection{Spectra with increasing $U'$}
\label{sec:uprimespec}

We now consider dynamics on the low-energy scales characteristic of the Kondo resonance that arises in the SC phase.
For convenience we shall analyse the two regions $0\le\tilde U'\le \tilde U$ and $\tilde U'\ge\tilde U$ separately. The former displays the crossover from $SU(2)$ to $SU(4)$ behaviour on increasing $\tilde U'$ towards the $SU(4)$ point $U'=U$, 
whereas in the latter region we show how the Kondo resonance is destroyed as $\tilde U'$ moves through the phase transition. Throughout the SC phase the Fermi level spectrum is (as above) pinned at the unitarity limit, 
$\pi \Gamma D_{i}(\omega =0) =1$; and we note that this behaviour is accurately
captured by the numerics on employing the standard $A_{\Lambda}$-factor 
($\Gamma \rightarrow A_{\Lambda}\Gamma$)~\cite{kww1}
that corrects for the finite-$\Lambda$ discretization
inherent to the NRG.

\begin{figure}
\centering\includegraphics{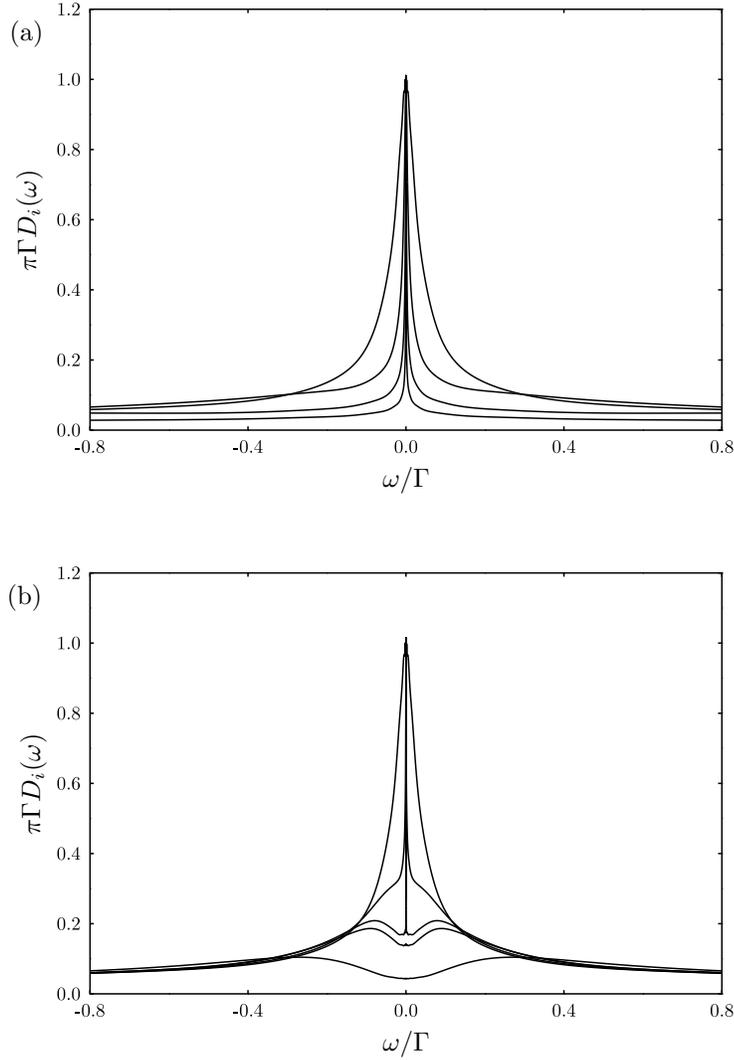}
\caption{\label{fig:specincup} Evolution of the single particle spectrum $\pi \Gamma D_i(\omega)$ vs. $\omega/\Gamma$ with increasing $\tilde U'$ at fixed $\tilde U=7$. (a) shows the crossover from $SU(2)$ to $SU(4)$ physics, $\tilde U'=0$, $6.5$, $6.9$ and $7$ (from bottom to top); while (b)  for
$\tilde U'=7$, $7.03$, $7.044$, $7.048$ and $7.1$ (from top to bottom)
shows the destruction of the Kondo resonance as the phase transition occurring at 
$\tilde{U}'_{c} \simeq 7.046$ is crossed.} 
\end{figure}

Figure~\ref{fig:specincup}(a) 
thus shows $\pi \Gamma D_i(\omega)$ versus $\omega/\Gamma$ for a fixed, strongly
correlated $\tilde U=7$, with increasing $\tilde U'=0$, $6.5$, $6.9$ and $7$
(in order of increasing resonance width). 
As expected from the thermodynamic results discussed in I, the $SU(2)$ physics of $\tilde U'=0$ is found to persist with increasing $\tilde U'$ until one gets very close to the $SU(4)$ point. For $\tilde U'=6.5$ for example, the functional form of the Kondo resonance is essentially the same as for $\tilde U'=0$, the only difference being that its width is increased slightly in proportion to the Kondo temperature 
$T_K$ (the evolution of $T_K$ itself being shown explicitly in figure 5 of I).

As $U'$ moves closer to $U$ however, the \lmfour{} fixed point begins to play an increasingly important r\^ole in shaping the low-frequency form of the spectrum. This is evident in the case $\tilde U'=6.9$, where there is a distinct shoulder in the Kondo resonance when $\omega/\Gamma$ is on the scale of the crossover from \lmfour to \lmtwo{} ($\sim (U-U')/\pi\Gamma=0.1$). We discuss this point in more detail when considering the scaling characteristics of the spectrum in \sref{sec:uniscale}; for now we simply note that at frequencies below the crossover scale the spectrum is again of the same $SU(2)$ form as for $\tilde U'=0$, whereas on higher scales the tails of the resonance approach those of the $\tilde U'=\tilde U$ $SU(4)$ spectrum
which is also shown in the figure.

Figure~\ref{fig:specincup}(b) illustrates the situation on the other side of the $SU(4)$ point: we show $\pi \Gamma D_i(\omega)$ versus $\omega/\Gamma$, again for fixed $\tilde U=7$ but now with $\tilde U'=7$, $7.03$, $7.044$, $7.048$ and $7.1$, spanning as such the transition occurring at $\tilde{U}'_{c} \simeq 7.046$. 
As shown in figure 5 of I, with increasing $\tilde{U}'$ the Kondo scale $T_K$ in the SC phase now decreases exponentially rapidly (following the Kosterlitz-Thouless form equation (6.3) of I), and ultimately vanishes at the critical $\tilde{U}'_c$. 
The vanishing of the low-energy Kondo scale as the transition is approached is seen vividly in the behaviour of the Kondo resonance. This narrows rapidly with
increasing $\tilde{U}'$ (figure~\ref{fig:specincup}b for
$\tilde U'=7.03$ and $7.044$), while maintaining the unitarity limit $\pi \Gamma D_{i}(\omega=0) = 1$. The Kondo resonance collapses `on the spot' at $\tilde{U}'=\tilde{U}'_{c}$, and for  $\tilde{U}' > \tilde{U}'_{c}$ in the CO phase leaves only an incoherent continuum around the Fermi level (with $\pi \Gamma D_{i}(0) \lesssim
0.2$).

\subsection{Universal scaling behaviour}
\label{sec:uniscale}
One of the most important results for the single-impurity Anderson model is 
the universality of the Kondo resonance, arising in the strongly correlated Kondo regime ($\tilde{U} \gg 1$) when the frequency axis is rescaled in terms of the Kondo scale $T_K$. Here, we examine corresponding issues of universality for the DQD model in the different regimes of the SC phase.

\begin{figure}
\centering\includegraphics{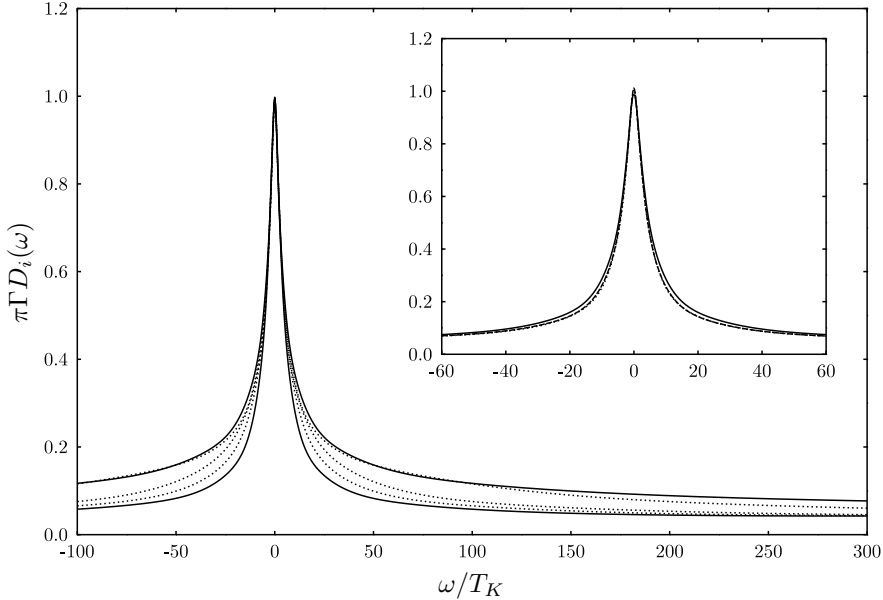}
\caption{\label{fig:scalspecsk}Scaling spectra in the spin-Kondo regime. Main figure: $\pi \Gamma D_i(\omega)$ vs. $\omega/T_K$ for fixed $\tilde U=7$ and $\tilde U'=0$ (top solid line), $6.95$, $6.98$, $6.99$ (dotted lines from top to bottom) and $7$ (bottom solid line). Inset: the collapse of the $SU(4)$ scaling spectrum onto a common form with increasing $\tilde U$, $\pi \Gamma D_i(\omega)$ vs. $\omega/T_K$ for $\tilde U'=\tilde U=7$ (solid line), $9$ (dashed line) and $11$ (dotted line). Note that the $\tilde U=9$ and $11$ spectra are virtually indistinguishable.}
\end{figure}

First recall briefly the scaling behaviour of the spectra $\pi \Gamma D_{i}(\omega)$ for $\tilde U'=0$, where
the single-impurity limit (at particle-hole symmetry) is recovered. For 
$\tilde U\gg 1$, spectra for different values of $\tilde U$ collapse onto a common form (see e.g.\ \cite{hewson}) when the frequency axis is rescaled as $\omega/T_K$,
with \emph{no} dependence on the `bare' parameters of the model.
Once this `scaling spectrum' is known, the spectrum for any particular $\tilde U$ 
can thus be determined on an absolute frequency scale simply from a knowledge of $T_K$ alone. The scaling spectrum naturally embodies the universal physics of the 
many-body Kondo resonance --- the high-energy, non-universal Hubbard satellites for example (at $|\omega| \simeq \case{U}{2}$) are obviously `projected out' of the scaling spectrum.
We show the $\tilde U'=0$ 
scaling spectrum, $\pi \Gamma D_i(\omega)$ versus $\omega/T_K$, as the uppermost solid line in figure~\ref{fig:scalspecsk}; and remark in passing that the analytical structure of it is both rich and non-trivial, see 
e.g.~\cite{nld}.

The $\tilde{U}'=0$ scaling spectrum pertains of course to the $SU(2)$ Kondo model.
One would clearly expect a similar scaling spectrum to exist at 
$\tilde U'= \tilde U$, in this case reflecting $SU(4)$ Kondo physics. That this is indeed so is seen in the
inset to figure~\ref{fig:scalspecsk}, which shows $\pi \Gamma D_i(\omega)$ versus 
$\omega/T_K$ for $\tilde U'=\tilde U=7$, $9$ and $11$. From this the collapse onto a universal form is immediately apparent. Of course, this $SU(4)$ scaling spectrum differs substantially from its $SU(2)$ counterpart, as shown in figure~\ref{fig:scalspecsk} where the $SU(4)$ spectrum is plotted as the lower solid line.

Having shown distinct scaling behaviour at the $SU(2)$ and $SU(4)$ points, an
obvious question arises: how do the scaling spectra evolve from $SU(2)$ to $SU(4)$ behaviour in the intervening region $0<\tilde U'<\tilde U$? As mentioned in
the discussion of section~\ref{sec:uprimespec}, it is in fact the non-universal 
scale $U-U'$ which controls this crossover; and the magnitude of $U-U'$ in relation to $T_K$ determines whether the scaling spectrum will be essentially that of $\tilde U'=0$, $\tilde U'=\tilde U$, or a combination of the two. For example, if $\tilde U'$ is not exponentially close to $\tilde U$, the ratio $(U-U')/T_K$ will be very large (recall
that $T_{K}$ is always exponentially small in strong coupling), the \lmfour{} fixed point will thus have basically no influence on the low-frequency dynamics, and the spectrum should therefore collapse onto the $SU(2)$ scaling spectrum.
Similarly, if $(U-U')\ll T_K$, then one would expect the scaling spectrum to be essentially unchanged from the $SU(4)$ spectrum on all but possibly the lowest frequency scales (and even here the Friedel sum rule guarantees that $\pi \Gamma D_{i}(\omega) \rightarrow 1$ as $\omega/T_{K} \rightarrow 0$ throughout the SC phase). When $(U-U')\sim {\cal{O}}(T_K)$ however, we do expect to see qualitatively different scaling spectra: here the crossover from $SU(2)$ to $SU(4)$ occurs on a scale of 
order $T_K$ itself, and we would thus expect the spectra to show characteristics of both regimes.

Our numerical results agree with the arguments given above. For $\tilde U'\lesssim 6.9$ and frequencies in the range shown in figure~\ref{fig:scalspecsk}, the single particle spectra plotted as a function of $\omega/T_K$ collapse essentially perfectly onto the $U'=0$ $SU(2)$ scaling spectrum. For larger values of $\tilde U'$ however, the ratio $(U'-U)/T_K$ moves into the frequency range under consideration and hence the spectra begin to show signs of $SU(4)$ tails. For example, the dotted lines in figure~\ref{fig:scalspecsk} show $\pi \Gamma D_i(\omega)$ versus $\omega/T_K$ for $\tilde U'=6.95$, $6.98$ and $6.99$, from which it is clear that with increasing $\tilde U'$ towards $\tilde{U}' =\tilde{U}$, the $SU(2)$ low-frequency behaviour holds over a  progressively decreasing range of $\omega/T_K$. For values of $\tilde U'$ much closer still to 
$\tilde U$, the numerically obtained scaling spectra are virtually indistinguishable from that of the $SU(4)$ line $U'=U$.

\begin{figure}
\centering\includegraphics{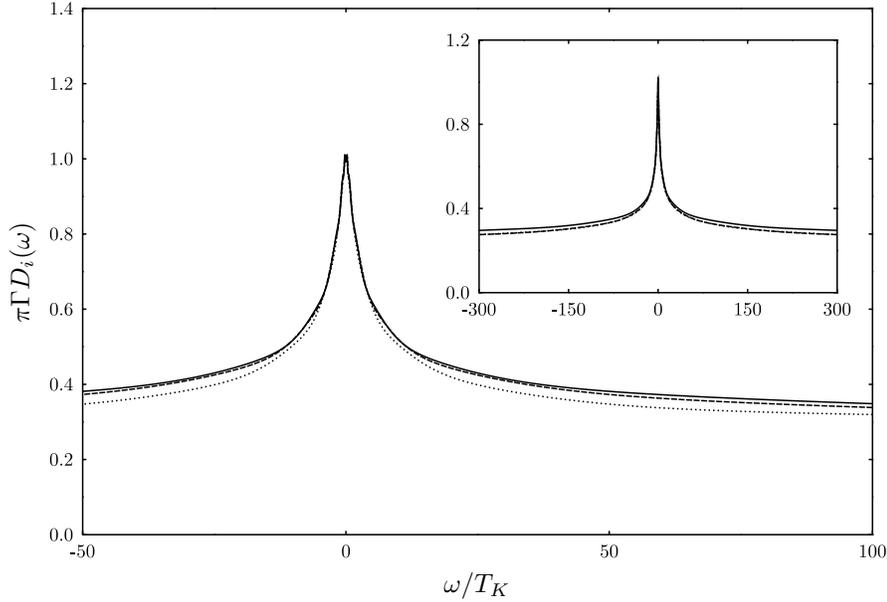}
\caption{\label{fig:scalspecck} Scaling spectra in the charge-Kondo regime. Main figure: $\pi \Gamma D_i(\omega)$ vs. $\omega/T_K$ for $\tilde U=7$ and $\tilde{U}'=7.03$ (dotted line), $7.04$ (dashed line) and $7.044$ (solid line). Inset: $\pi \Gamma D_i(\omega)$ vs. $\omega/T_K$ for $\tilde U=7$ (solid line), $8$ (dashed line) and $9$ (dotted line) where $(\tilde{U}'-\tilde{U})/(\tilde{U}'_c-\tilde{U})$ is fixed at $\simeq 0.86$. Note that the spectra for the largest two values of $\tilde{U}$ are essentially indistinguishable.}
\end{figure}

The discussion above has focussed on the crossover from $SU(2)$ to $SU(4)$ behaviour, i.e. for $\tilde U'\le \tilde U$. We now show that there is a \emph{third} scaling spectrum in the SC phase, which describes the behaviour in the charge-Kondo regime 
($\tilde U<\tilde U'<\tilde U'_c$) as the transition is approached and the Kondo
scale $T_{K}$ acquires its asymptotic Kosterlitz-Thouless form (equation (6.3) of I,
vanishing exponentially rapidly as $\tilde{U}' \rightarrow \tilde{U}'_{c}-$).
The main part of figure \ref{fig:scalspecck}
 shows $\pi \Gamma D_i(\omega)$ versus $\omega/T_K$ for fixed $\tilde U=7$ and 
$\tilde{U}'=7.03$, $7.04$ and $7.044$: as $\tilde{U}'$ approaches 
$\tilde{U}'_c$ ($\simeq 7.046$), the Kondo temperature $T_K$ rapidly diminishes and the rescaled spectra collapse onto a common form. The scaling spectrum shows a distinct `charge-Kondo' resonance, the tails of which tend not to zero as $\omega/T_K\to\infty$, but to the value 
$\sin^2 [\case{\pi}{2}(1-\case{1}{\sqrt{2}})]\simeq 0.2$ which is the zero-frequency value of the spectrum on entering the CO phase (section \ref{sec:potscat}).

In the inset of \fref{fig:scalspecck}, we show further that this scaling spectrum is independent of the value of $\tilde U$ (for strong-coupling $\tilde U\gg 1$), the same spectrum being obtained close to the phase boundary for $\tilde U=7$, $8$ and 
$9$. For each $\tilde{U}$, the value of $\tilde{U}'$ has been chosen sufficiently 
close to the phase boundary that the Kondo scale $T_{K}$ is well described by its asymptotic Kosterlitz-Thouless form (see e.g.\ figure 5 of I); the
$\tilde{U}'$s being chosen in practice such that
$(\tilde{U}'-\tilde{U})/(\tilde{U}'_c-\tilde{U}) = 0.86$.

To summarise, we have shown that the scaling behaviour of the spectra throughout the SC phase can for the most part be understood quite simply in terms of the three scaling spectra obtained in the limits $\tilde U'=0$, $\tilde U'=\tilde U$ and $\tilde U'=\tilde U'_c-$. The $\tilde U'=0$ scaling spectrum in fact captures the behaviour in the vast majority of the phase, being the appropriate description for all $U-U'\gg T_K$. As one moves closer to the $SU(4)$ line, the spectra at low-frequencies cross over to the $SU(4)$ scaling spectrum shown in figure~\ref{fig:scalspecsk}, and then beyond $\tilde U'=\tilde U$ the spectra rapidly collapses onto the charge-Kondo scaling spectrum shown in figure~\ref{fig:scalspecck}.

\section{Discussion}

  Finally, we comment briefly on symmetry breaking perturbations which destroy
the stability of the CO fixed point. 
Particle-hole asymmetry does not of course fall into this category. 
As we have emphasised several times here and in I the particle-hole symmetric case
($\epsilon = -\case{U}{2}-U'$) is entirely representative, and the 
essential physics robust to departure from it.
However the model we have considered is `left/right' symmetric: equivalent
dots (e.g.\ $\epsilon_{R} = \epsilon_{L} \equiv \epsilon$), and equivalent coupling to the conduction channels/leads ($\Gamma_{L} =\Gamma_{R} \equiv \Gamma$).
If $L/R$ symmetry is broken it is readily shown that while the SC fixed point remains stable, additional relevant perturbations arise for the CO fixed point
and hence render it unstable. Flows in the vicinity of the CO fixed point thus ultimately flow away from it under renormalization, and cross over instead to the SC fixed point.

  Since the quantum phase transition between the SC and CO phases is strictly destroyed by breaking $L/R$ symmetry, the question arises: do visible remnants 
of the transition nonetheless remain? Indeed they do. To illustrate this we consider explicitly the
case of detuning the dot levels e.g.\ by applying a different gate voltage to 
each dot, taking $\epsilon_{L,R} = \epsilon \pm \case{1}{2}\delta\epsilon$ (with 
$\epsilon = -(\case{U}{2}+U')$ as usual). Here one expects intuitively that the 
crossover scale for flows away from the CO fixed point will be 
determined by $\delta\epsilon$; and that if $\delta\epsilon$ is not large compared 
to the $SU(4)$ scale $\tkfour$, then the erstwhile transition will be `smeared out' 
but still in essence visible as a clear crossover. That this is so is illustrated 
in figure 6, where we show NRG results for the differential conductance 
$\pi \Gamma D_{L}(0)$ \emph{vs} $\tilde{U}'$ for three different $\delta\epsilon$,
viz $\delta\epsilon/\tkfour = 0, 0.3$ and $2$ (with $\tkfour/\Gamma \simeq 0.01$
the $\tilde{U}'=\tilde{U} = 7$ $SU(4)$ scale for $\delta\epsilon =0$). The
pristine transition occurring for $\delta\epsilon =0$ is as anticipated smeared out,
on a scale determined by $\delta\epsilon$; albeit remaining clearly visible 
(note that the lowest $\tilde{U}'$ shown in figure 6 is $\tilde{U}'=6.9$),
and as such robust in an obvious sense. 

\begin{figure}
\centering\includegraphics{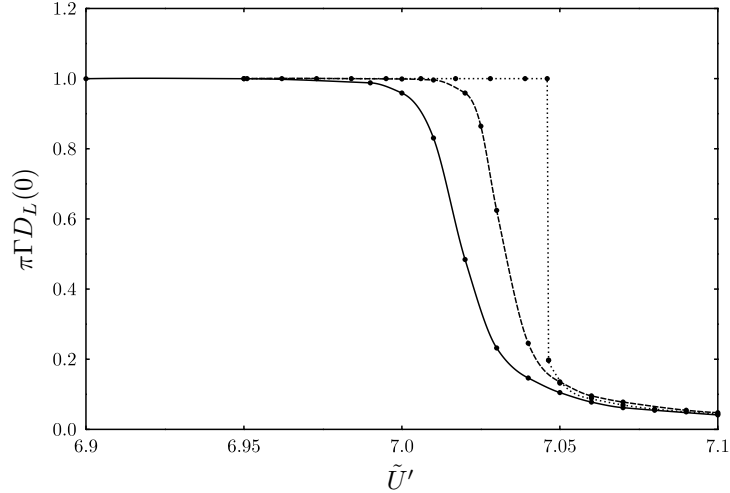}
\caption{\label{fig:deltaeps} Linear differential conductance $\pi \Gamma D_{L}(\omega=0)$ \emph{vs} $\tilde{U}'$ for a fixed $\tilde{U} =7$ and $\delta\epsilon/\tkfour = 0$ (dotted line), $0.3$ (dashed) and $2$ (solid). Note that the lowest
$\tilde{U}'$ shown in the figure is $\tilde{U}'=6.9$.}
\end{figure}

\ack 

We are grateful to the EPSRC for supporting this research, and to the Royal Society for travel support.


\section*{References}
\begin{thebibliography}{99}
\bibitem{I} Galpin M R, Logan D E and Krishnamurthy H R 2006 \emph{J. Phys.: Condens. 
Matter}, submitted for publication.
\bibitem{wilson} Wilson K G 1975 \emph{Rev. Mod. Phys.} \bf 47 \rm 773
\bibitem{kww1} Krishnamurthy H R, Wilkins J W and Wilson K G 1980 \emph{Phys. Rev. B} \bf 21 \rm 1003 
\bibitem{kww2} Krishnamurthy H R, Wilkins J W and Wilson K G 1980 \emph{Phys. Rev. B} \bf 21 \rm 1044
\bibitem{hewson} Hewson A C 1993 \emph{ The Kondo Problem to Heavy Fermions} (Cambridge: Cambridge University Press)
\bibitem{ddprl}For a preliminary account of some of these results see: Galpin M R, Logan D E and Krishnamurthy H R 2005 \emph{Phys. Rev. Lett.} \bf 94 \rm 186406
\bibitem{meirwin} Meir Y and Wingreen N S  1992 \emph{Phys. Rev. Lett.} \bf 68 \rm 2512 
\bibitem{hewsoncosti} Costi T A, Hewson A C and Zlati\'c V 1994 \emph{J. Phys.: Condens. Matter} \bf 6 \rm 2519
\bibitem{garst} Garst M \emph{et al} 2004 \emph{Phys. Rev. B} \bf 69 \rm 214413
\bibitem{HofSch} Hofstetter W and Schoeller H 2002 \emph{Phys. Rev. Lett.} \bf 88 
\rm 016803
\bibitem{JPCM} Galpin M R and Logan D E 2005 \emph{J. Phys.: Condens. Matter}
\bf 17 \rm 6959
\bibitem{lma1} Logan D E, Eastwood M A and Tusch M T 1998 \emph{J. Phys.: Condens. Matter} \bf 10 \rm 2673
\bibitem{lmamtg} Glossop M T and Logan D E 2002 \emph{J. Phys.: Condens. Matter}
\bf 14 \rm 6737
\bibitem{lmasg1} Logan D E and Glossop M T 2000 \emph{J. Phys.: Condens. Matter}
\bf 12 \rm 985
\bibitem{nld} Dickens N L and Logan D E 2001 \emph{J. Phys.: Condens. Matter}
\bf 13 \rm 4505


\end{thebibliography}
\end{document}